\documentclass[letterpaper,11pt,fleqn]{article}
\usepackage{bm,graphicx,amsmath,appendix,amsfonts,amsthm,amssymb,color,geometry,setspace,enumerate, aeguill,array,multirow,xcolor,subfig,epstopdf,natbib,parskip,float,lscape,booktabs,enumitem,tabularx}
\usepackage[nottoc,numbib]{tocbibind}
\usepackage[]{hyperref}
\usepackage[font=small,labelfont=bf,singlelinecheck=false]{caption}
\setlist[itemize]{leftmargin=*}
\usepackage{mathpazo}
\usepackage[T1]{fontenc}
\allowdisplaybreaks

\title{{\bf A time-varying finance-led model for U.S. business cycles}\thanks{I would like to thank Rudiger von Arnim for several discussions on earlier versions of this paper. The usual disclaimer applies.}}
\author{Marcio Santetti\thanks{Economics Department, Skidmore College. Email: \color{blue}\texttt{msantetti@skidmore.edu}.}}

\begin{document}

\maketitle

\begin{abstract}
\noindent This paper empirically assesses predictions of Goodwin's model of cyclical growth regarding demand and distributive regimes when integrating the real and financial sectors. In addition, it evaluates how financial and employment shocks affect the labor market and monetary policy variables over six different U.S. business-cycle peaks. It identifies a parsimonious Time-Varying Vector Autoregressive model with Stochastic Volatility (TVP-VAR-SV) with the labor share of income, the employment rate, residential investment, and the interest rate spread as endogenous variables. Using Bayesian inference methods, key results suggest (\textit{i}) a combination of profit-led demand and profit-squeeze distribution; (\textit{ii}) weakening of these regimes during the Great Moderation; and (\textit{iii}) significant connections between the standard Goodwinian variables and residential investment as well as term spreads. Findings presented here broadly conform to the transition to increasingly deregulated financial and labor markets initiated in the 1980s.

\end{abstract}

\bigskip


\bigskip


\newpage

\section{Introduction}\label{sec:intro}

The "Goodwin pattern" describes cyclical interactions between economic activity and the labor share of income. From both theoretical and empirical viewpoints, a substantial body of research documents a counter-clockwise movement between these two measures for the post-war U.S. economy. Furthermore, diverse econometric estimations with high-frequency data support the demand and distributive regimes experienced by the country as profit-led and profit-squeeze, respectively. 

Despite providing a coherent view of cyclical movements of growth and distribution, empirical contributions in this literature rarely extend economic activity to capital accumulation and interactions between the real and financial sectors. Several theoretical traditions agree on residential (housing) investment as the leading factor setting off a period of economic recovery, but its integration in models of demand and distribution with financial extensions is still limited.

Furthermore, the larger body of applied time-series models that aim to estimate the associations between growth and distribution over time relies on standard Vector Autoregressive (VAR) models, which provide a single set of point estimates that are supposed to describe an entire sample period. Even if one tests for and applies structural breaks, this methodology will still deliver unique parameters, which may not capture continuous changes in the data. In the case of U.S. economy, its post-war era is marked by crises and recoveries of different natures and lengths. More specifically, it has undergone two broad periods: a "Golden Age" of accelerated growth until the stagflation of the 1970s, and a "neoliberal era" featuring monetary tightening and increasingly deregulated financial and labor markets. These latter measures comprise the "Great Moderation" of the mid-1980s. This way, econometric models that aim to better explain reality should take potential endogenous and exogenous continuous fluctuations into account.

A recent effort on this subject has been made by \cite{maldniki22}. Focusing solely on the demand regime experienced by the U.S. economy over the 1947--2019 interval, the authors estimate a Time-Varying VAR (TVP-VAR) model with the rate of output growth and the labor share as activity and distributive variables, respectively. This baseline model and its variants find support for continuous changes in the demand regime, which has become less profit-led over the last forty years.\footnote{The authors test other specifications, including the debt-to-disposable income ratio, labor productivity, public deficit, and trade balance as third variables, as well as replacing output growth with the rate of capacity utilization.}

On the distributive regime, \cite{setterfield23} claims that the profit-squeeze mechanism has been weakening in the U.S. economy. As a consequence of shifting institutional and economic policy conditions over the last decades, the working class has become less capable of bargaining for higher shares of aggregate income in their favor, and thus periods of labor market tightening have not been sufficient to produce sustained increases in the labor share. The author, however, does not put this claim to econometric scrutiny.

A similar claim---this time grounded on empirical estimations---has been made by \cite{goldstein99}. The author estimates a baseline two-variable VAR model with the profit share and the employment rate, inferring that the profit-squeeze mechanism has been weakening since the 1970s, and with more vigor after the Great Moderation policies. Furthermore, the study also estimates a three-dimensional model including capital accumulation as a state variable---namely nonresidential investment flows---, arriving at similar conclusions.\footnote{The analysis carried out by \cite{goldstein99} builds on the "cyclical profit squeeze" theory of business cycles presented in \cite{bocrotty}. See also \cite{basu13} and Section \ref{sec:lit} in this paper.} 

Finally, capital accumulation is the channel through which the real sector connects to the financial side of the aggregate economy. This link has been the object of several theoretical studies \citep{ryoo, taylor12, sordi14, stockmichell}. From an empirical standpoint, \cite{rezai13} and \citet[section~5.1]{barrales21} propose recursive VAR models with financial extensions. Their estimations find support for the profit-led/profit-squeeze pattern. These, do not, however, include a measure of investment as a covariate, and do not consider possible continuous changes in the regime over time.

The present study tackles the dynamics of demand and distributive regimes with their financial interactions over time in a single, comprehensive empirical model. Through an aggregative approach,\footnote{\cite{blecker16} distinguishes between "structural" and "aggregative" models for estimating demand regimes. While the first category decomposes aggregate demand into its segments---such as investment, consumption, and the trade balance---, the second does not rely on decompositions. For a "structural" approach to this issue, see, for instance, \cite{stamegna22}.} it puts forth a parsimonious, recursive Time-Varying VAR model with Stochastic Volatility (TVP-VAR-SV) with the labor share of income, the employment rate, residential investment, and the interest rate spread as endogenous variables. This way, aggregate demand and income distribution are integrated with the financial side, with residential investment being the channel between the goods and money markets. Furthermore, the additional assumption of stochastic volatility allows for better capturing time variations in the endogenous variables over time, as the premise of homoskedasticity may bias time-varying coefficients.

To preface results and conclusions, the paper finds statistical evidence in favor of the Goodwin pattern. First, through descriptive statistics, it shows that, over six different business cycles, a counter-clockwise movement is the recurring pattern between activity and distribution. Second, lead-lag analysis provides evidence of activity leading the labor share, finance leading residential investment, and the latter leading the business cycle. 

The model finds support for a weakening of both profit-led demand and profit-squeeze distribution until 2019, when activity and distributive variables show more vigorous responses to labor share and employment structural shocks, respectively. Furthermore, the labor market reacts positively to residential investment shocks, while the interest rate spread responds negatively to employment disturbances. Finally, residential investment reacts positively to financial shocks, in line with theoretical contributions that couple real and financial sectors in models of growth and distribution. All analyzed responses to exogenous shocks show a decreased variance after the business-cycle peak of the 1980s, in 1981Q3, until the pre-Great Recession peak, in 2007Q4. This finding is consistent with the reduction in macroeconomic volatility during the Great Moderation years \citep{davis08}.

The paper proceeds as follows. Section \ref{sec:lit} briefly reviews the literature on demand and distributive regimes, with emphasis on empirical studies. It also explores how financial factors integrate with economic activity and income distribution, with the main channel being residential investment. Section \ref{sec:data} provides descriptive statistics of the selected variables over different U.S. business cycles. Section \ref{sec:methodology} outlines the empirical methodology, based on a time-varying VAR model with stochastic volatility. Section \ref{sec:results} explores results, and Section \ref{sec:concl} concludes.

\section{Related Literature}\label{sec:lit}

The "Goodwin pattern" is characterized by counter-clockwise fluctuations in a two-dimensional plane where an economic activity variable lies on the horizontal, and a distributive measure on the vertical axis. The former may be proxied by the output gap, the rate of capacity utilization, and the employment rate, among others. The latter is usually proxied by the wage (labor) share of aggregate income.\footnote{The distributive variable may also be proxied by the profit share, and this would generate clockwise cycles in an activity-distribution plane.} At business-cycle frequencies, this has been a consistent pattern across different capitalist countries, including the U.S. economy. 

Building on the groundbreaking work by \cite{goodwin82}, numerous contributions have provided theoretical variations of the "canonical" model \citep{skott89, barbosa06, stockmichell, rada21}. The same applies for empirical studies \citep{rezai13, carv16, barrales21, mendieta22, gcri}. Despite their particular research questions and different econometric techniques, two key points are pervasive in these studies: the regimes governing demand and distribution over time.

First introduced by \cite{bha90}, the two standard demand regimes of an economy are wage- and profit-led. Profit-led demand features an increase in aggregate investment (and exports, in an open-economy scenario) following an increase (decrease) in the profit (wage) share of income. This upswing in investment compensates for a reduction in aggregate consumption, as the share accrued by workers diminishes. On the other hand, an economy is wage-led if rising consumption compensates for a fall in investment due to an increase (decrease) in the wage (profit) share.

From a distributive aspect, there are also two leading patterns. The first is a profit squeeze, described by a tighter labor market pushing for an increasing wage share. The second is a wage squeeze, where increases in economic activity lead to income distribution toward profits, so the present level of investment is held constant. The center of these dynamics lies in the labor market, whose state influences the struggle between working and capitalist classes in their conflicting claims over income shares \citep{skott89}. As institutional and social norms may better reflect the interests of the class with greater bargaining power in a given historical moment, distributive variables are usually assumed to be exogenous in empirical models \citep{blecker19}.

A combination of profit-led demand and profit-squeeze distribution (PL/PS, for short) has been the most consistent finding supported by empirical studies for the U.S. economy \citep{barrales21}. These results are mostly derived from Vector Autoregressive (VAR) models, where identification strategies play a crucial role. Both in a recursive case, where an ordering is imposed upon the variables, and in structural models, where time restrictions are placed with no particular variable ordering, proper theoretical motivations must guide the statistical analysis. 

In \cite{goodwin82}, the central stage of economic dynamics is the labor market. Activity is represented by the employment rate, with the labor share as the distributive measure. The goods market is assumed to be in equilibrium, and higher employment rates imply greater workers' ability to bargain for a higher wage rate and, consequently, a higher labor share. In line with Classical/Keynesian approaches, the wage share converges more slowly to its steady state than the employment rate. As the labor share is the ratio between real wages and labor productivity, even if the latter may act procyclically and relatively fast, real wages will only recover with a time lag. When this is properly accounted for in VAR settings, the PL/PS pattern is robust to diverse variable definitions, data sources, and model dimensionalities.

Although focusing on the real side of the economy, the standard Goodwin model does not comprise a more detailed and independent role of aggregate investment in its narrative. Consequently, it does not explore how financial conditions affect capital accumulation and its outcomes on activity and distribution. Therefore, the interplay between real and financial sectors is central for a broader assessment of business cycles. Across both standard and alternative theoretical approaches, residential (housing) expenditures play the role of leading other activity variables and, thus, the beginning of an expansion \citep{davis05, kydland, barbosa08, fiebiger18}. 

The inclusion of capital accumulation as a state variable, along with other activity and distributive measures, has precedents in the applied literature. \cite{goldstein99} estimates a baseline VAR model with the employment rate and the profit share, as well as a three-dimensional model with nonresidential expenditures, finding evidence in support of a weakening of the profit-squeeze regime especially after the mid-1980s. \cite{basu13} build on the latter study by estimating a VAR model with filtered unemployment rate, profit share, and nonresidential investment covariates, inferring that the profit-squeeze mechanism is present both in the regulated and neoliberal eras of the U.S. economy. Lastly, \cite{basugautham} consider nonresidential investment flows along with the labor share, the exchange rate, labor productivity growth, the unemployment rate, and the utilization rate in a VAR setting. Focusing on the impact of exogenous shocks to the wage share,  the authors find support for the PL/PS regime in the U.S. economy between 1973 and 2018.

The aforementioned studies, however, do not consider aggregate investment nor its residential component to analyze business cycles. \cite{gcri} is a recent effort on this issue, presenting empirical estimates that residential investment leads the business cycle, while nonresidential expenditures lag it. Moreover, the Goodwin pattern is consistent with the inclusion of aggregate and disaggregated investment variables for the U.S. economy's post-war period. The authors, nonetheless, do not include financial interactions in their structural VAR models.

Monetary extensions in models of cyclical growth is not a new agenda. \cite{foley87} explores liquidity-profit rate cycles, focusing on firms' capital expenditure and borrowing decisions setting off a new cycle. In a model consisting of firms' debt as the sum of money and other financial assets (all relative to the capital stock) and the profit rate, the latter proxies for the activity variable. The author argues that growing investment and sales raise profit rates, thus leading to a greater supply of loanable funds and further accumulation. However, if investment grows faster than the money supply, there will be a shortage of liquidity and, consequently, macroeconomic instability.\footnote{This model is further extended by \cite{araujo20} with the inclusion of a nominal wage Phillips curve, and the authors' unstable equilibrium outcome is only overcome by state intervention through monetary policy. See also \cite{proano06} and \cite{proano11} for a similar conclusion.} \cite{taylor12} also explores real-financial cycles in a model with the accumulation rate as the activity variable and Tobin's $q$ as a measure of finance. The main prediction is a negative effect of activity on finance, and a halt in the cycle led by increasing asset prices. In both studies, income distribution is assumed constant. On the other hand, \cite{barrales21} infer that both models predict a clockwise cycle in an activity-finance plane.

An integration between the goods and money markets is not complete without considering financial institutions. On this issue, diverse perspectives investigate the channels through which the latter may influence real activity. The literature on this matter is vast, and a thorough survey is beyond the scope of this paper. However, given the research questions at hand, works along New Keynesian and Minskyan lines are worth mentioning, as the role of agents' expectations are fundamental to understand risk management, credit provision, and their effects on the real sector.

\cite{ashin10} present a critique of New Keynesian models concerning the importance of financial institutions vis-\`a-vis economic activity. The authors define intermediaries' leverage costs as a function of risk and short-run interest rates, while expected profitability is proxied by interest rate (term) spreads. Furthermore, variations in credit supply depend on how these agents manage their balance sheets. By determining short-term rates, monetary policy movements thus influence intermediaries' profitability expectations and the slack in their assets and liabilities.\footnote{This view is directly opposed to other works in this literature, which do not consider an active role of financial institutions and their connection to real activity. For instance, \cite{bern04} and \cite{svensson04} claim that movements in short-term interest rates derived from monetary policy are only relevant  to determine long-run interest rates, stock prices, and exchange rates. This way, banks are a passive channel through which the central bank transmits monetary policy, with no direct effect in the real sector.} A lowering of target short-term rates increases the term spread, thus expanding financial intermediaries' balance sheets. As this triggers more optimistic expectations on future profitability, a greater willingness to supply loanable funds impacts interest-sensitive activities, such as residential investment and durable goods consumption.
 
Another strand of the Classical/Keynesian literature coupling real and financial sectors derives from including the financial instability hypothesis put forth verbally in \cite{minsky75,minsky86} and more formally within the Minskyan tradition \citep{charles08, fazzari08}. According to this view, capitalist economies are inherently unstable due to the instability of investment. Business cycles would arise from the interactions of demand in the goods market and financial fragility, which may be proxied by the debt-to-profit ratio of non-financial firms.

Several works in this vein connect Goodwinian and Minskyan features by exploring firms' and financial intermediaries' decisions throughout the business cycle. \cite{stockmichell} set up a two-variable model in financial instability and demand. The former is a positive function of the latter, while demand is an inverse function of fragility. An economic recovery bolster firms' optimistic perspectives on investment, leading to a higher willingness to accumulate debt. By sharing a similar belief, banks also engage in greater lending. Then, at higher levels of output, firms' and banks' balance sheets endogenously deteriorate, leading to a collapse. The paper, however, does provide a clear definition of the "fragility" variable. 

On the negative feedback from financial fragility to aggregate demand, \cite{stockmichell} point out two potential explanations. First, central banks may respond to an increased price level by raising interest rates following an expansion period. Second, a decision to increase interest rates may come from individual banks, as these observe rising firms' debt ratios. Regardless of the nature of increasing borrowing costs, this leads to a fall in investment and a halt to the business cycle. Finally, the authors predict a counter-clockwise cycle in an activity-financial instability plane. \cite{keen95} also combines Goodwin's real side analysis with financial instability through a dynamic system formed by the unemployment rate, the labor share, and the debt ratio. Building on similar priors, \cite{sordi14} include an independent investment function as a fourth state variable, thus allowing for disequilibrium in the goods market. In other words, the authors allow aggregate investment to differ from savings during an upswing in economic activity.\footnote{See also \cite{sportelli95} on the use of an independent investment function in a Goodwinian setting.}

The aforementioned works on real-financial connections do not pursue empirical estimations, and applied contributions on this issue are still scarce. A few exceptions are the recursive VAR estimations of \cite{rezai13}, who finds support for the PL/PS regime in the U.S. by including the interest rate spread and inflation in a model \`a-la \cite{barbosa06}, and \citet[section~5.1]{barrales21}, who include a measure of Tobin's $q$ along with the output gap and the labor share, also finding evidence of profit-led demand and profit-squeeze distribution.

Regardless of including financial factors or solely focusing on the real sector, one fundamental limitation emerges across all works which use vector autoregressions in this literature. The usual procedure returns single point estimates that are supposed to describe the entire sample period the researcher is interested in. Even if one tests for an applies structural breaks, splitting the sample into two or more periods will still return single sets of point estimates, without allowing for continuous changes in the variables and their co-movements over time.\footnote{See, for instance, \cite{mendieta22}.} 

The issue above is of special relevance for the U.S. economy, whose post-war period can be (broadly) divided into two main periods: before and after the "Great Moderation" of the 1980s. The first is marked by a "Golden Age" period of Keynesian policy orientation and accelerated growth followed by the stagflation of the 1970s, and the introduction of contractionary monetary policies with the goal of reducing the volatility of economic activity \citep{davis08}. The second period, beginning in the mid-1980s, features increasing deregulation in financial and labor markets, and rising income inequality. The Great Recession of 2007/08 motivated discussions of secular stagnation, as the growth rate of output has declined in the U.S. and other developed economies \citep{limunoz}.

The first empirical effort to consider time-varying features in models of growth and distribution is \cite{maldniki22}. Employing a Time-Varying Parameter VAR (TVP-VAR) approach,  the authors estimate several models to assess possible changes in the demand regime experienced by the U.S. economy over the 1947--2019 period. Their main premise is to employ a statistical methodology that allows for continuous changes in the parameters over time, without the need for structural breaks and/or splitting the sample period to evaluate their differences. Overall, the main findings include a strengthening of the profit-led regime from the post-World War II years until the 1970s, followed by a decline over the last four decades. 

Without relying on econometrics, \cite{setterfield23} argues that the profit-squeeze regime has been weakening in the U.S. over the last decades. Building on the institutional view of \cite{corn}, the author poses that the rise of neoliberal policies, deunionization, and business outsourcing throughout the years following the Great Moderation have institutionalized working class insecurity. Its direct consequence is a weaker bargaining position vis-\`a-vis capitalists, resulting in subsequent failures to push income distribution in their favor. This fact brings about a breakdown of the Goodwin pattern. Lastly, the author remarks that whether this will become a definitive regularity or a temporary event remains an answered question. As previously mentioned, \cite{goldstein99} finds empirical support for a weakening of the distributive regime, especially over the Great Moderation years.

The present study tackles the demand regime inquiry of \cite{maldniki22}, and expands the research question to other two main issues. First, the claim made by \cite{goldstein99} and \cite{setterfield23} regarding a weakening of the profit-squeeze mechanism; second, it investigates the connection between real and financial sectors, illustrating how employment, investment, and finance shocks affect activity and financial variables. It also adopts a time-varying VAR approach, including stochastic volatility, in order to capture continuous changes in these variables and potential changes in the variance of exogenous shocks across different U.S. business cycles. This latter assumption is particularly appealing in the context of changing labor market and financial institutions put forth by the Great Moderation years, whose main goal was to reduce the buoyancy of economic activity, in order to control inflation and induce growth.

Before any further econometric work, Section \ref{sec:data} explores some statistical regularities involving activity, distributive, and financial variables over the studied sample period. Beyond complementing the theoretical issues outlined in this section, this analysis will provide the empirical basis to properly identify the model presented in Section \ref{sec:methodology}.

\section{Data and Descriptive Statistics}\label{sec:data}

This section's objective is twofold. Firstly, it presents the cyclical interactions involving activity, distributive, and financial variables in the post-war U.S. macroeconomy. Secondly, it provides empirical justifications that, in addition to the literature outlined in the previous section, will help to properly motivate the modeling strategy in Section \ref{sec:methodology}.

The present analysis uses the following data:

\begin{itemize}
	\item Labor (wage) share of income, $\psi_t$: calculated from the Bureau of Economic Analysis' (BEA) National Income and Product Accounts (NIPA), Table 1.10. It is the ratio between total compensation (including public employment, line 2) and the sum of the latter and net interest (line 11), rental income (line 14), corporate profits (line 15), and capital consumption (line 21);
	\item Employment rate, $e_t$: calculated as 100 minus the seasonally adjusted civilian unemployment rate (UNRATE) series, available at the Federal Reserve Economic Database (FRED);
	\item Residential investment, $g_t$: accessible as an index (2012=100) from the NIPA, Table 1.1.3;\footnote{Both $g_t$ and $\psi_t$ were log-transformed.}
	\item Interest rate spread, $s_t$: computed as the difference between the market yield on U.S. Treasury securities at 10-year maturity rate (FRED GS10 series) and the 3-month Treasury bills secondary market rate (FRED TB3MS series). 
\end{itemize}

The choice of financial variable is not trivial. The main requirement concerns its interactions with real activity at business-cycle frequencies. This way, variables such as household debt, net worth, and real estate prices move on lower frequencies than those explored here \citep{barbosa08, taylor12}. The interest rate spread is then chosen due to its informative content and its predictive power over future residential investment and, consequently, the employment rate. Proxy variables for interest rate spreads vary in the literature, and this study follows the measure in works such as \cite{estrella91} and \cite{adrian19}, with 10-year yields reflecting the return on longer-term investments and 3-month Treasury yields echoing monetary policy movements.\footnote{For instance, \cite{friedman98} use the difference between commercial paper and Treasury bill rates as a spread measure, while \cite{laubach03} adopt the spread between current real interest rates and own computations of a natural interest rate.}

This study adopts the view that the channel connecting the real and financial sectors is residential investment. Furthermore, the latter is the key driver of the business cycle according to several strands of the literature. And if housing investment is the cycle's driver, financial institutions are its engine, through the provision of loanable funds. Therefore, the latter agents play an effective role in developing the expectations that will frame different phases of the business cycle.

An active feedback between financial institutions and real activity is put forth by \cite{ashin10}. According to the authors, intermediaries influence the goods market by determining the price of risk, and fluctuations in the supply of credit follow from how these agents manage their balance sheets. In their model, the authors proxy financial institutions' expected profitability with interest rate spreads, and changes in monetary policy directly affect expected profits and how much slack banks have in their assets and liabilities. For instance, a lowering of target short-term rates may boost expectations, expanding intermediaries' balance sheets. As financial institutions borrow from others at shorter terms, but supply credit to non-financial units at longer terms, a larger term spread implies a larger willingness to borrow. Consequently, interest-sensitive demand components, such as residential investment, are directly affected by how intermediaries manage their expectations and, ultimately, their balance sheets.\footnote{Additional views on the interactions between financial institutions and real activity are provided in \citet[ch.~4]{semmler10}, \cite{chinnkucko}, and \cite{adrian19}.}

For robustness purposes, additional measures of the labor share and finance were used. For the former, the Bureau of Labor Statistics' (BLS) "headline" measure for the non-farm business sector; for the latter, a measure of Tobin's $q$, calculated as the ratio (in basis points) between the value of the non-financial corporate sector’s equities (FRED MVEONWMVBSNNCB series) and its net worth (TNWMVBSNNCB series). As the empirical results are consistent when using these different variables, robustness checks will not be explicitly reported for the sake of compactness.

All variables are in quarterly frequency, with the sample period ranging from 1953Q2 to 2022Q4. To focus on high-frequency dynamics, the paper concentrates on the variables' cyclical component, that is, on the respective deviations from their long-run trends. Among several different trend-cycle decomposition options, the chosen method is the Hamilton filter \citep{hamilton}, as it avoids end-of-sample bias intrinsic to other techniques, such as the Hodrick-Prescott filter \citep{hp97}.\footnote{The Hamilton filter was specified with the default lookahead and lag periods of 8 and 4 quarters, respectively. In addition, its use shrinks the full sample period to 1956Q1--2022Q4.}

The descriptive analysis begins with two-dimensional, time-connected scatter diagrams. By visualizing how two different variables are associated over time, the orientation of the relationships may help to better grasp potential cyclical movements between them.  For more clarity, the sample period is split into six different peak-to-peak business cycles, as defined by the National Bureau of Economic Research (NBER): 1969Q4--1973Q4, 1973Q4--1981Q3, 1981Q3--1990Q3, 1990Q3--2001Q1, 2001Q1--2007Q4, and 2007Q4--2019Q4. Finally, following a similar approach to \cite{zip11}, one additional year (4 quarters) is added to the second peak of each sub-period, in case the cycle completes after the official dating.\footnote{One key difference to \cite{zip11} is that the authors use the Hodrick-Prescott filter to decompose their series.}

\begin{figure}[H]
  \includegraphics[width=\textwidth]{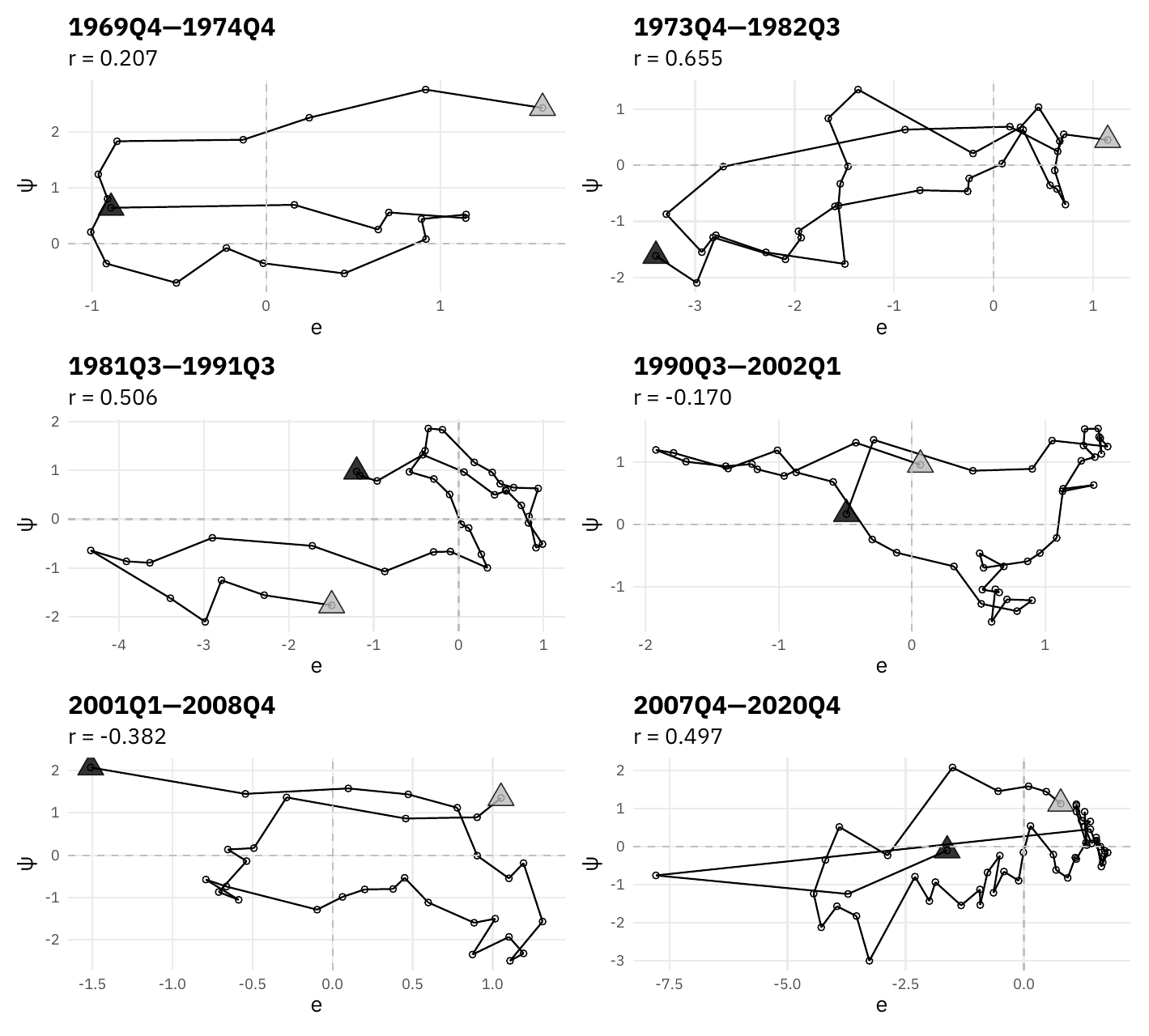}
  \caption{Time-connected scatter diagram, $(\psi, e)$ plane. Gray and black triangles denote the first and last quarter, respectively.}
  \label{fig:e-psigr-cyc}
\end{figure}

The relationship between the employment rate (\textit{x}-axis) and the labor share (\textit{y}-axis) is presented in Figure \ref{fig:e-psigr-cyc}, with gray and black triangles denoting the sub-period's first and last quarters, respectively. In addition, all graphs show the cycle's Pearson correlation coefficient ($r$) between the variables. All panels show clear counter-clockwise cycles. Except for a brief clockwise movement from 1986 until 1991, this behavior corroborates the "Goodwin pattern," as described and tested in several works \citep{barbosa06, barrales21, gcri}.

Furthermore, for two consecutive sub-periods (1990Q3--2002Q1, 2001Q1--2008Q4), the variables show negative correlations, and the highest coefficient is observed in the cycle preceding the Great Moderation. The most recent sub-period exhibits a major recovery in this linear association, with the caution that the COVID-19 first quarters show sharp decreases in both $\psi$ and $e$.

\bigskip

\begin{center}
	[FIGURE \ref{fig:e-psigr-cyc} ABOUT HERE]
\end{center}

\bigskip

\begin{figure}[H]
  \includegraphics[width=\textwidth]{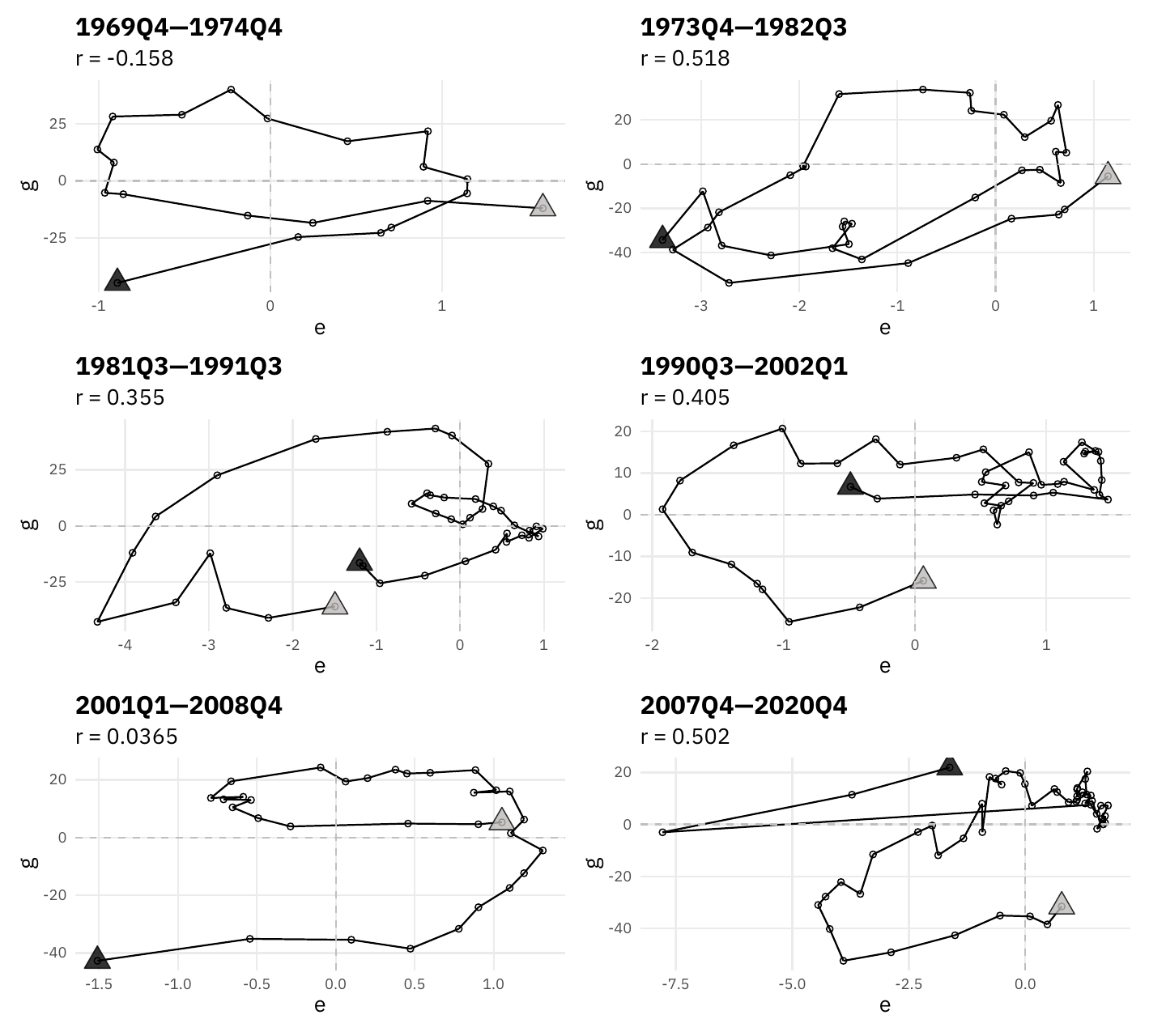}
  \caption{Time-connected scatter diagram, $(e, g)$ plane. Gray and black triangles denote the first and last quarter, respectively.}
  \label{fig:e-gr-cyc}
\end{figure}

Figure \ref{fig:e-gr-cyc} illustrates the joint behavior of the two activity variables: residential investment (\textit{y}-axis) and the employment rate (\textit{x}-axis). The first cycle shows a negative correlation, while the 2001Q1--2008Q4 subsample has two periods of stagnant investment's and major drops in employment's cyclical behaviors. Finally, the highest correlation is observed in the second sub-period, similarly to the previous figure.

\bigskip

\begin{center}
	[FIGURE \ref{fig:e-gr-cyc} ABOUT HERE]
\end{center}

\bigskip

\begin{figure}[H]
  \includegraphics[width=\textwidth]{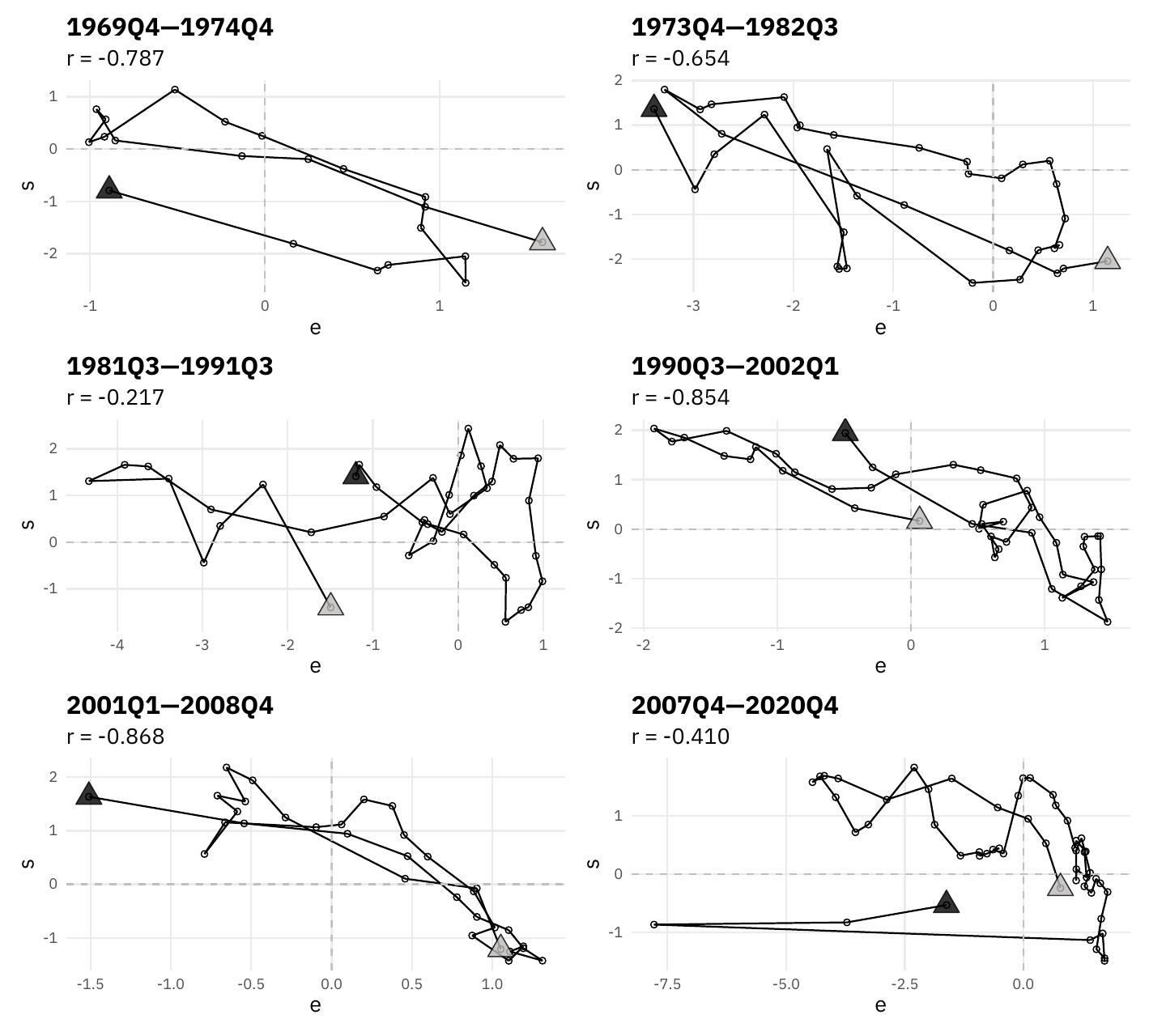}
  \caption{Time-connected scatter diagram, $(e, s)$ plane. Gray and black triangles denote the first and last quarter, respectively.}
  \label{fig:e-q-cyc}
\end{figure}

\begin{figure}[H]
  \includegraphics[width=\textwidth]{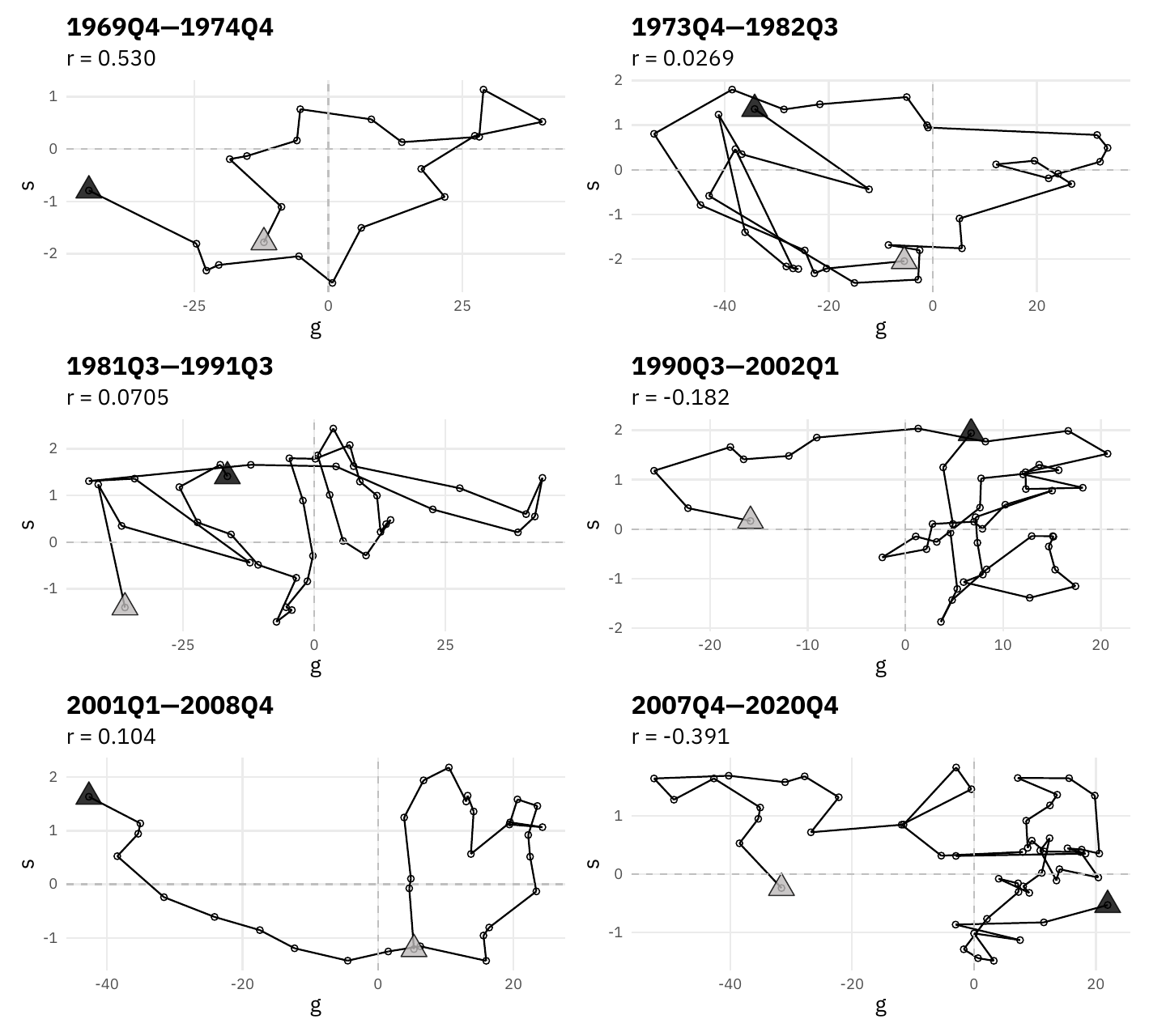}
  \caption{Time-connected scatter diagram, $(g, s)$ plane. Gray and black triangles denote the first and last quarter, respectively.}
  \label{fig:gr-q-cyc}
\end{figure}

Lastly, Figures \ref{fig:e-q-cyc} and \ref{fig:gr-q-cyc} present the dynamic associations between finance (\textit{y}-axis) and the the two activity variables. For both $(e, s)$ and $(g, s)$ planes, all sub-periods show a clockwise pattern. It is also worth noting that the association between the employment rate and finance decreases between the first and third cycles, then recovering and reaching its highest levels from the 1990s until the Great Recession. On the other hand, while residential investment and the rate spread also show a decreasing linear relationship in the same first cycles, from the 1990s until the most recent data the correlation coefficient oscillates in sign.

\bigskip

\begin{center}
	[FIGURES \ref{fig:e-q-cyc} AND \ref{fig:gr-q-cyc} ABOUT HERE]
\end{center}

\bigskip

In spite of its informative content, the previous visualizations and statistics are not sufficient to establish lead/lag relationships among these variables. As this feature will be relevant for the identification strategy in the next section, computing cross-correlations allows searching for potential leading and lagging relationships between the variables examined here. Using notation from \citet[ch. 1]{shumwaystoffer}, the cross-correlation coefficient between $x$ and $y$ at lag \textit{h}, $\hat{\rho}_{xy}$, can be defined by

\begin{equation}
	\hat{\rho}_{xy} = \dfrac{\hat{\gamma}_{xy}(h)}{\sqrt{\hat{\gamma}_{x}(0)\hat{\gamma}_{y}(0)}},
\end{equation}

where $\hat{\gamma}_{xy}(h)$ is the sample cross-covariance between time series $x_t$ and $y_t$ at lag \textit{h}; and the denominator is the square-rooted contemporaneous interaction between the autocorrelations of $x_t$ and $y_t$.

\begin{figure}[H]
  \includegraphics[width=\textwidth]{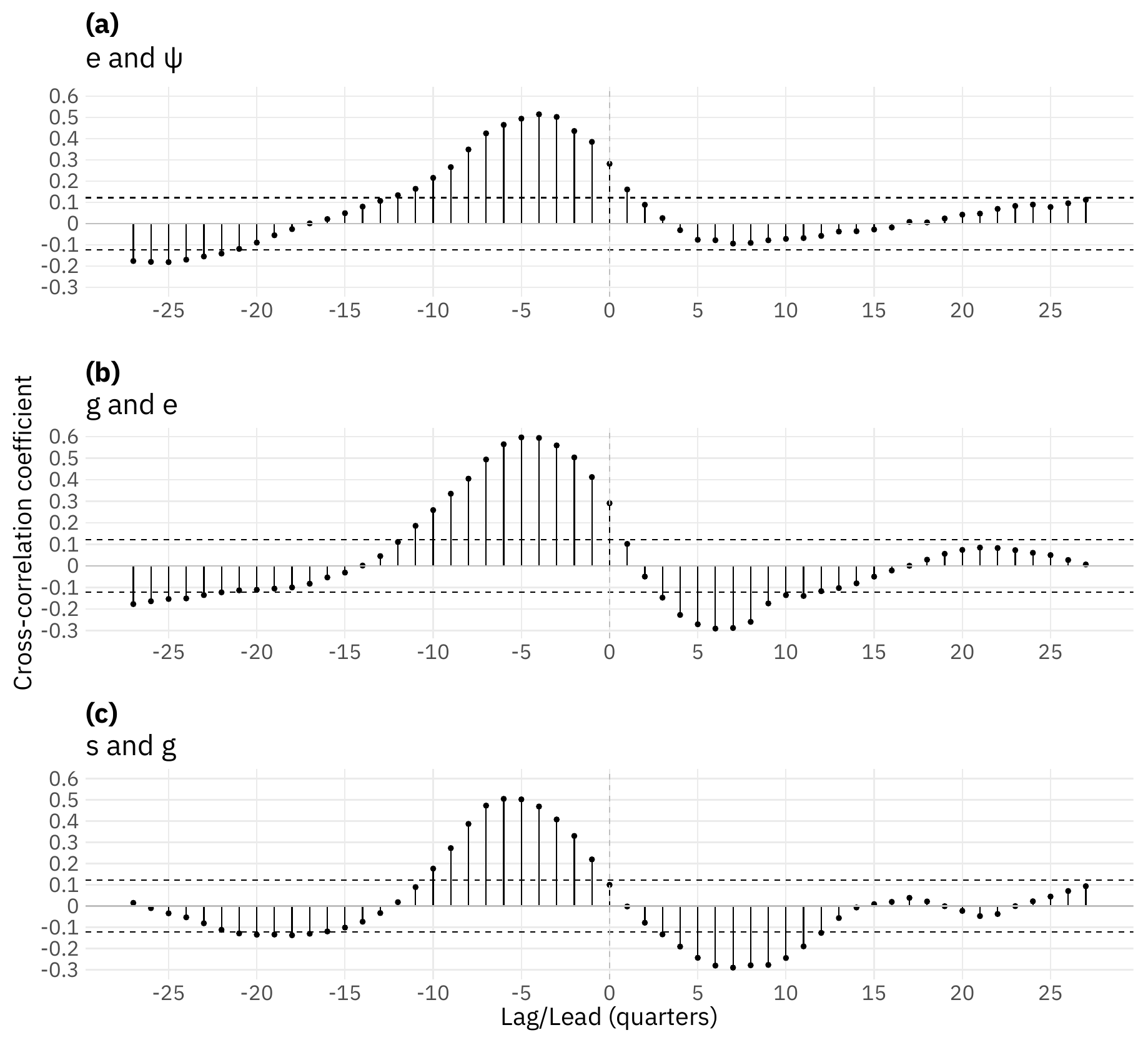}
  \caption{Cross-correlation coefficient plots. \small Panel (a): employment rate \textit{vs.} labor share; panel (b): residential investment \textit{vs.} employment rate; panel (c): interest rate spread \textit{vs.} residential investment. Dashed horizontal lines indicate $\pm 2 \times 1/\sqrt{n}$, with $n = 268$ observations.}
  \label{fig:ccf}
\end{figure}

Figure \ref{fig:ccf} displays the cross-correlation coefficients (\textit{y}-axis) at different lags (negative) and leads (positive, \textit{x}-axis) for all relevant relationships.\footnote{As only the cyclical components of the time series are being studied, using the Hamilton filter constitutes a "prewhitening" (i.e., de-trending) step to properly analyze lead/lag relationships \citep{shumwaystoffer}.}  With the highest cross-correlation coefficients being observed in lagging quarters, panels (a)--(c) suggest that ($i$) the employment rate leads the labor share; ($ii$) residential investment leads the employment rate; and ($iii$) finance leads residential investment. These results conform to both theoretical priors and empirical evidence analyzed in Section \ref{sec:lit}.

\bigskip

\begin{center}
	[FIGURE \ref{fig:ccf} ABOUT HERE]
\end{center}

\bigskip

This exploratory section concludes with two descriptive tables. Firstly, Table \ref{tab:tab1} presents the standard deviations (volatilities) of each series' cyclical components by cycle. Over time, only the two activity variables showed an increase in their volatilities, while the labor share and finance became less spread out. Finally, the interest rate spread is overall the least volatile variable, and residential investment is the most buoyant. 

\bigskip

\begin{center}
	[TABLE \ref{tab:tab1} ABOUT HERE]
\end{center}

\bigskip

As a complement to the previous table, Table \ref{tab:tab2} reports the standard deviations of each variable by merging cycles in two distinct periods: the post-war (1956Q1--1984Q4) and the neoliberal (1985Q1--2019Q4) eras, denoting pre- and post-Great Moderation phases.\footnote{This nomenclature was given in \cite{mendieta22}. In addition, this study also follows the sample period split adopted by the authors, as they found statistical evidence of a structural break in their data.} The first period captures the last 17 years of the U.S. economy's "Golden Age," followed by the crises of the 1970s and early 1980s. The second, on the other hand, spans over the period after the "Volcker shock." As pointed out by different examples in the literature on U.S. business cycles, the second half of the 1980s is marked by changing monetary policy and labor market institutional compositions, leading to a reduced  volatility of macroeconomic shocks and economic growth \citep{stock02, fogli, mendieta22}. Regarding the variables discussed in this study, residential investment and the interest rate spread show reduced standard deviations, while the labor share and the employment rate have increased volatilities.

\bigskip

\begin{center}
	[TABLE \ref{tab:tab2} ABOUT HERE]
\end{center}

\bigskip

This sample period can be summarized by changing dynamics in the co-movements of distributive, activity, and financial variables in the U.S. macroeconomy. Although most of the empirical phase trajectories studied here display consistent cyclical orientations, these happen in different regions of the two-dimensional planes, as well as with different completion lengths across cycles. Furthermore, the volatilities of each variable are not constant over the cycles, and these diagnostics cannot be ignored in empirical estimations that aim to better reflect reality. The next section addresses these issues.

\section{Empirical Methodology}\label{sec:methodology}

This section outlines the econometric technique and the identification strategy adopted in this study. First, it details the time-varying VAR methodology, and then proceeds to the recursive ordering specifying the model.

\subsection{Estimation technique}

Vector autoregressive (VAR) models are a widely used technique in empirical macroeconomics whenever the researcher is interested in the dynamic interactions shared by a set of two or more variables over time \citep{sims80, chris99}. However, one key drawback of standard VAR analysis is that it delivers a single set of parameters that are supposed to describe the entire sample period, thus not reflecting potential changes in the underlying structure of an economy.

The development of time-varying parameter VAR (TVP-VAR) models addresses this issue. The major contribution to this literature was made by \cite{pri05}, proposing a VAR methodology whose (\textit{i}) coefficients are allowed to vary over time and (\textit{ii}) variance need not be constant.\footnote{Despite being the most prominent contribution, \cite{pri05} was preceded by other time-varying VAR studies, such as \cite{cog01} and \cite{sims01}.}
 Regarding the first point, it assumes that each parameter follows a first-order random walk process. This specification allows for both temporary and permanent changes in the parameters. This assumption, however, does not avoid possible coefficient bias, as the variance of the model residuals can also shift over time. Thus, the addition of stochastic volatility (SV) drifts away from the standard Ordinary Least Squares (OLS) homoskedasticity assumption, and a random variability of innovations is also allowed \citep{nak}.

A time-varying parameter VAR model with stochastic volatility (TVP-VAR-SV) can be presented in reduced-form as follows:

\begin{equation}\label{eq:eq1}
	\mathbf{y}_t = X_t\bm{\beta}_t + A^{-1}_t\Sigma_t\varepsilon_t, \hspace{1cm} t = s+1,..., n,
\end{equation}

where $\mathbf{y}_t$ is a $k \times 1$ vector of endogenous variables; $\bm{\beta}_t$ is a $(k^2s \times 1)$ vector of time-varying coefficients; $X_t = I_k \otimes (\mathbf{y}_{t-1},..., \mathbf{y}_{t-s})$ is the Kronecker product of an identity matrix and the lagged observations of the endogenous variables; $A_t$ is a lower-triangular matrix of time-varying contemporaneous effects $a_{jt}$ of one-standard deviation shocks (innovations); $\Sigma_t$ is a diagonal matrix of time-varying standard deviations; and $\varepsilon_t$ s a vector of (additive) heteroskedastic structural shocks. Stochastic volatility arises when assuming that the logarithm of $\varepsilon_t$'s standard deviation elements, $h_{jt}$, also follow a first-order random walk process.

Given this setting, the parameters from Equation \ref{eq:eq1} can be defined by the following random walk regimes:

\begin{align}
\bm{\beta}_{t+1} = \bm{\beta}_t + u_{\bm{\beta}_t} \\
\mathbf{a}_{t+1} = \mathbf{a}_t + u_{\mathbf{a}_t} \\
\mathbf{h}_{t+1} = \mathbf{h}_t + u_{\mathbf{h}_t},
\end{align}

which describe possible time-varying behavior in the contemporaneous relations, lag structure, and the log-volatility of unexplained co-movements in the VAR model, respectively. Lastly, innovations from the reduced-form model $(\varepsilon_t)$, contemporaneous relationships $(u_{\bm{\beta}_t})$, lag structure $(u_{\mathbf{a}_t})$, and stochastic volatility $(u_{\mathbf{h}_t})$ are assumed to independent and identically follow a jointly Normal distribution.

A time-varying VAR model with stochastic volatility can be efficiently estimated in a Bayesian design. Following Bayes' theorem, if one is interested in an unknown parameter (or vector of parameters, in this case) $\bm{\theta}$, and has some available data $\mathbf{y}$, the former can be estimated conditional on the latter by

\begin{equation}\label{eq:bayes}
	p(\bm{\theta}|\mathbf{y}) = \dfrac{f(\mathbf{y}|\bm{\theta})p(\bm{\theta})}{\int f(\mathbf{y}|\bm{\theta})p(\bm{\theta})d\bm{\theta}} \ \ ,	
\end{equation}

where the left-hand side term is known as the posterior density of $\bm{\theta}$; the right-hand side's numerator is the product of the likelihood function for the data vector $\mathbf{y}$ and any prior information for the unknown parameter $\bm{\theta}$; and the denominator is the normalizing constant. As both the denominator and the likelihood function are usually intractable in TVP-VAR-SV contexts, Bayesian inference makes use of Markov Chain Monte Carlo (MCMC) algorithms to recursively sample from the posterior distribution.\footnote{For a detailed overview of TVP-VAR-SV models and MCMC algorithms, see \cite{nak} and \cite{krueger15}. Furthermore, these references detail the prior probability distributions assumed for the model parameters. Here, no change has been made to these prior distributions.}

\subsection{Identification}

The baseline model consists of the vector of endogenous variables $\mathbf{y}_t = (\psi_t, e_t, g_t, s_t)'$. Its identification follows the (recursive) Cholesky decomposition $\psi_t \rightarrow e_t \rightarrow g_t \rightarrow s_t$, with arrows indicating a contemporaneous effect from the left- to the right-side variable. In words, finance leads residential investment, which leads the other activity variable, the employment rate. The latter, finally, leads the labor share. In this setting, finance is the fastest-adjusting variable, while the labor share is the slowest.

This identification strategy follows directly from both theory and empirical evidence. Firstly, the distributive variable being exogenous to an activity measure (that is, with no contemporaneous relationship) is consistent with Classical/Keynesian approaches, as income distribution is determined by institutional factors, exogenous to the model. Secondly, both standard and alternative macroeconomic approaches agree on residential investment as the leading determinant of the business cycle, here represented by $e_t$. Thirdly, the interest rate spread is determined in financial markets, and provides a measure of finance that interacts with investment at higher frequencies. Fourthly, Section \ref{sec:data} provided the descriptive basis for this dynamic ordering, as well as for the assumption of stochastic volatility, given that the variance of each variable is not constant across different business cycles.

This paper follows the estimation procedure adopted by \cite{pri05}, with the MCMC algorithm further corrected by \cite{del15}. In order to estimate the posterior distributions of the parameters of interest, the model was run based on 55,000 draws, with the first 5,000 being discarded and serving as "burn-in" steps. The chosen lag order was 1, based on the Schwartz information criterion.

\section{Results}\label{sec:results}

This section explores results from the estimated TVP-VAR-SV model based on Impulse-Response Functions (IRFs). These consist of assessing the impact of (usually) a one-standard deviation shock to one endogenous variable on another over a specified time horizon. As the MCMC procedure yields not only one point estimate, but an entire distribution of the parameters of interest at each quarter, a compact and informative visual presentation of IRFs becomes a challenge. That said, IRFs will be displayed as follows:

\begin{itemize}
	\item Subfigures for six NBER business cycle peaks: 1973Q1, 1981Q3, 1990Q3, 2001Q1, 2007Q4, and 2019Q4;\footnote{IRFs have also been estimated for NBER troughs over the same sample period, with no change in overall results and conclusions.}
	\item Horizon divided in five different quarters ahead: 1, 4, 8, 12, and 20;
	\item Responses consisting of their entire posterior densities at each aforementioned quarter ahead, highlighting its posterior median and an underlying 66\% shaded density region.	This way, results incorporate key posterior estimates and their latent uncertainty, in line with Bayesian inference.
\end{itemize}

\begin{figure}[H]
  \includegraphics[width=\textwidth]{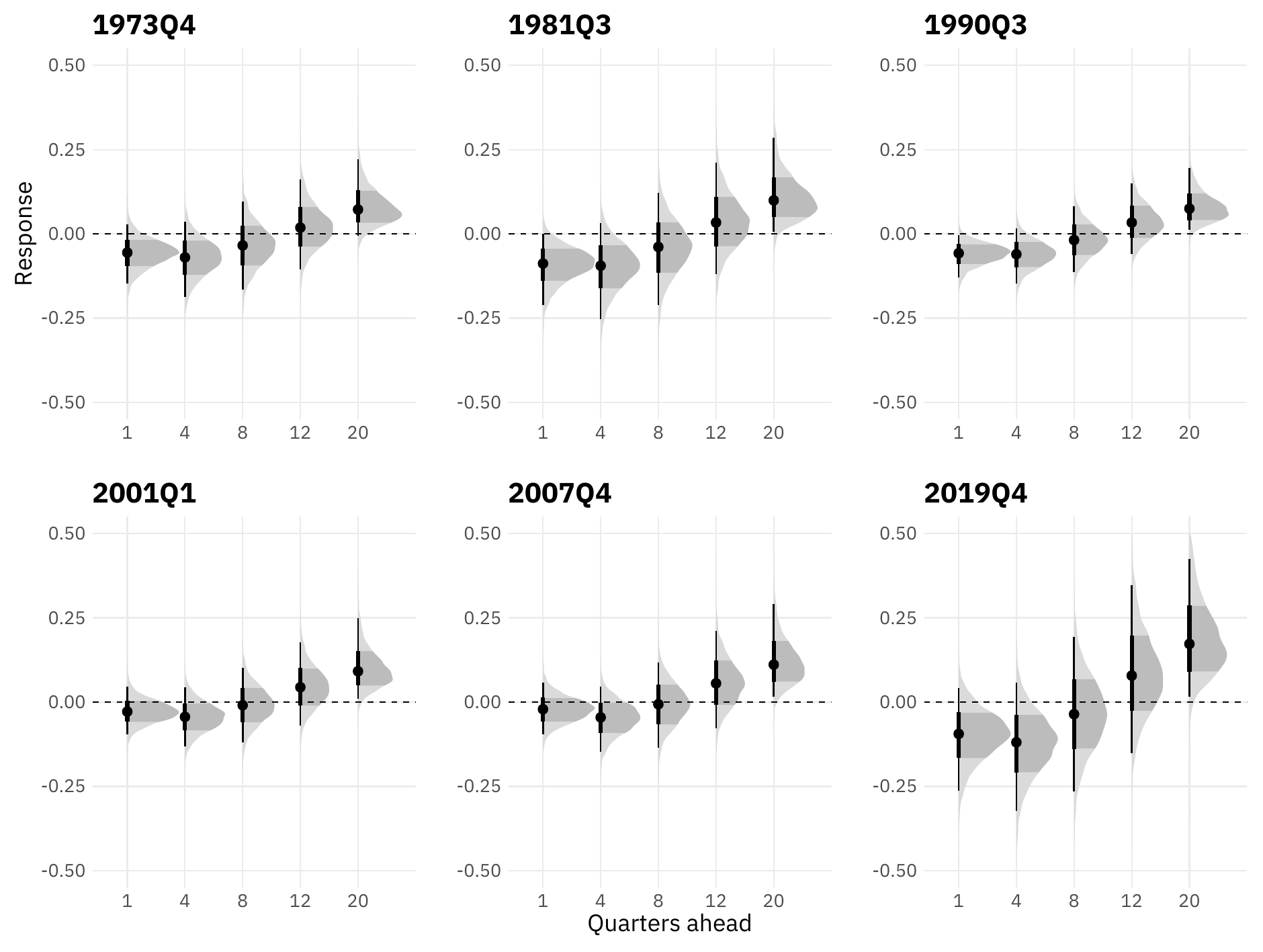}
  \caption{Posterior responses of the employment rate $(e_t)$ to a labor share $(\psi_t)$ shock at different NBER business cycle peaks. Black dots denote posterior medians, and darker shaded areas the 66\% posterior density region.}
  \label{fig:e-psigr-irf}
\end{figure}

Figure \ref{fig:e-psigr-irf} displays posterior responses of the employment rate to a wage share shock at different business cycle peaks. All subfigures exhibit a negative feedback over the initial quarters ahead (approximately two years), thus indicating a profit-led demand regime. The median responses then change in sign until the last quarter, indicating a weakening of profit-led demand at further horizons. How this sign change relates with the empirical phase trajectories shown in Section \ref{sec:data} is still an open question for further research. Furthermore, the three cycle peaks following the Great Moderation feature lower volatilities (i.e., narrower posterior distributions), if compared to the two first periods. Finally, the last cycle presents the highest posterior volatilities among all selected peaks.

These results are in line with the findings of \cite{maldniki22}. Although through a different visualization method, both time-varying VAR approaches find support for a weakening of the profit-led mechanism over the last decades.\footnote{The authors display IRFs by combining plots of different time horizons (0, 1, 3, 6, 9, and 12 quarters ahead) to evaluate structural changes in the variables, while here different business-cycle peaks are selected to observe their consistency over time.} By including the assumption of stochastic volatility, the present analysis also adds evidence of reduced buoyancy in the regime after the Great Moderation, with this pattern changing in the most recent business-cycle peak.

\bigskip

\begin{center}
	[FIGURE \ref{fig:e-psigr-irf} ABOUT HERE]
\end{center}

\bigskip

\begin{figure}[H]
  \includegraphics[width=\textwidth]{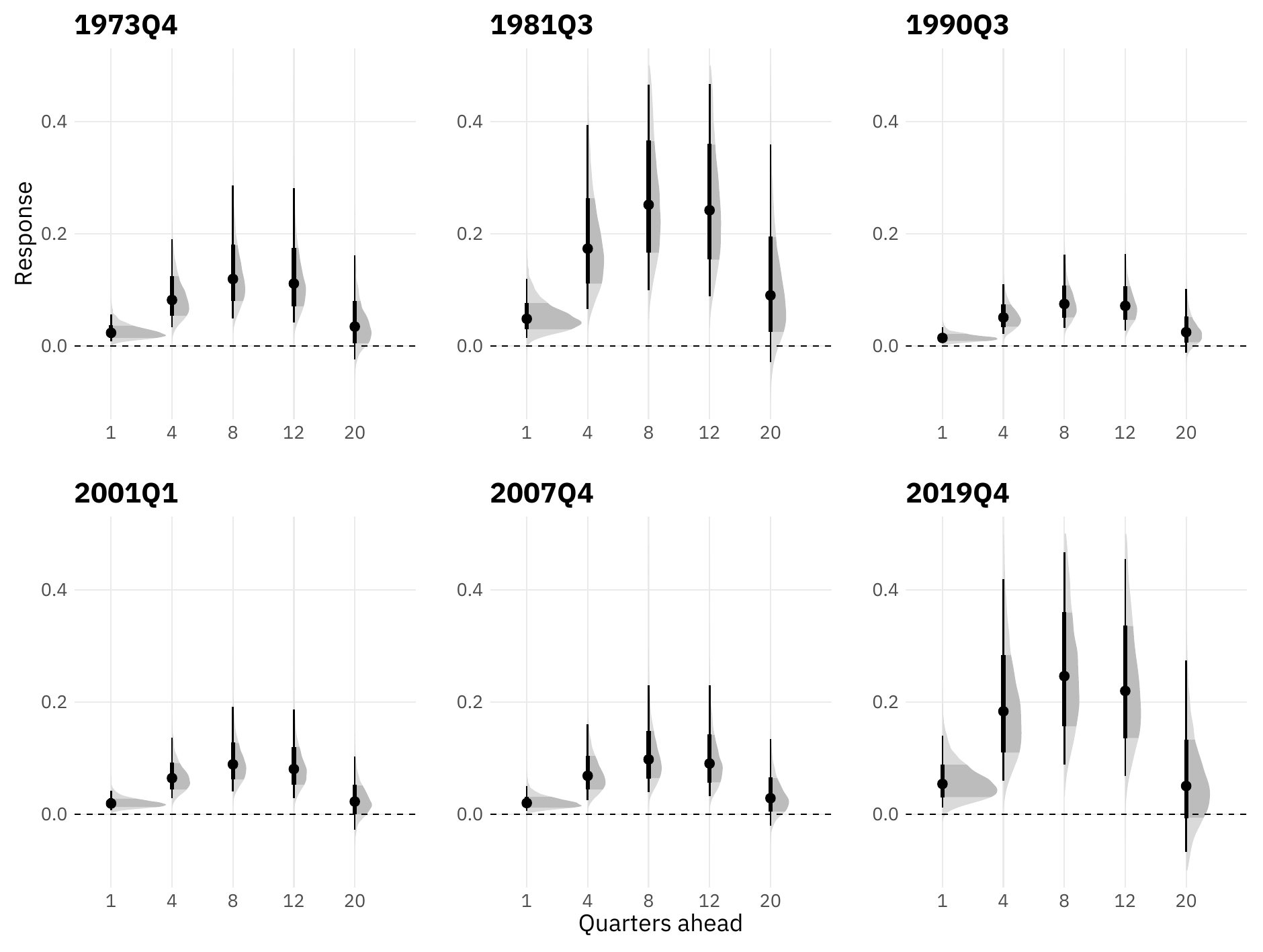}
  \caption{Posterior responses of the labor share $(\psi_t)$ to an employment rate $(e_t)$ shock at different NBER business cycle peaks. Black dots denote posterior medians, and darker shaded areas the 66\% posterior density region.}
  \label{fig:psigr-e-irf}
\end{figure}

In order to empirically assess the distributive regime for the U.S. economy, one can present how the labor share responds to an upswing in the labor market. This is shown in Figure \ref{fig:psigr-e-irf}. In all selected quarters, the wage share starts to respond more positively to labor market tightening between the second and fourth after-shock years, with the effect vanishing in longer periods. It is interesting to note that peaks with higher labor market activity (before the Great Moderation and pre-COVID periods) show higher volatility relative to the other sub-periods.

These results conform to a profit-squeeze regime. Estimations support a weakening of posterior responses over time, particularly in the three cycles after the Great Moderation \citep{goldstein99}. Similarly to the last figure, the distributive regime also features reduced buoyancy following the institutional changes of the mid-1980s. This behavior, however, changes in the last cycle, when the distributive variable's posterior responses resemble those prior to the Great Moderation. Referring to the hypothesis posed by \cite{setterfield23}, there is indeed evidence for a weakening of the profit-squeeze mechanism over the years, but with a recovery in the most recent cycle. The COVID-19 pandemic poses a relevant question mark on the continuation of this readjustment, and only when more data points become available this issue can be properly tackled again.

\bigskip

\begin{center}
	[FIGURE \ref{fig:psigr-e-irf} ABOUT HERE]
\end{center}

\bigskip

\begin{figure}[H]
  \includegraphics[width=\textwidth]{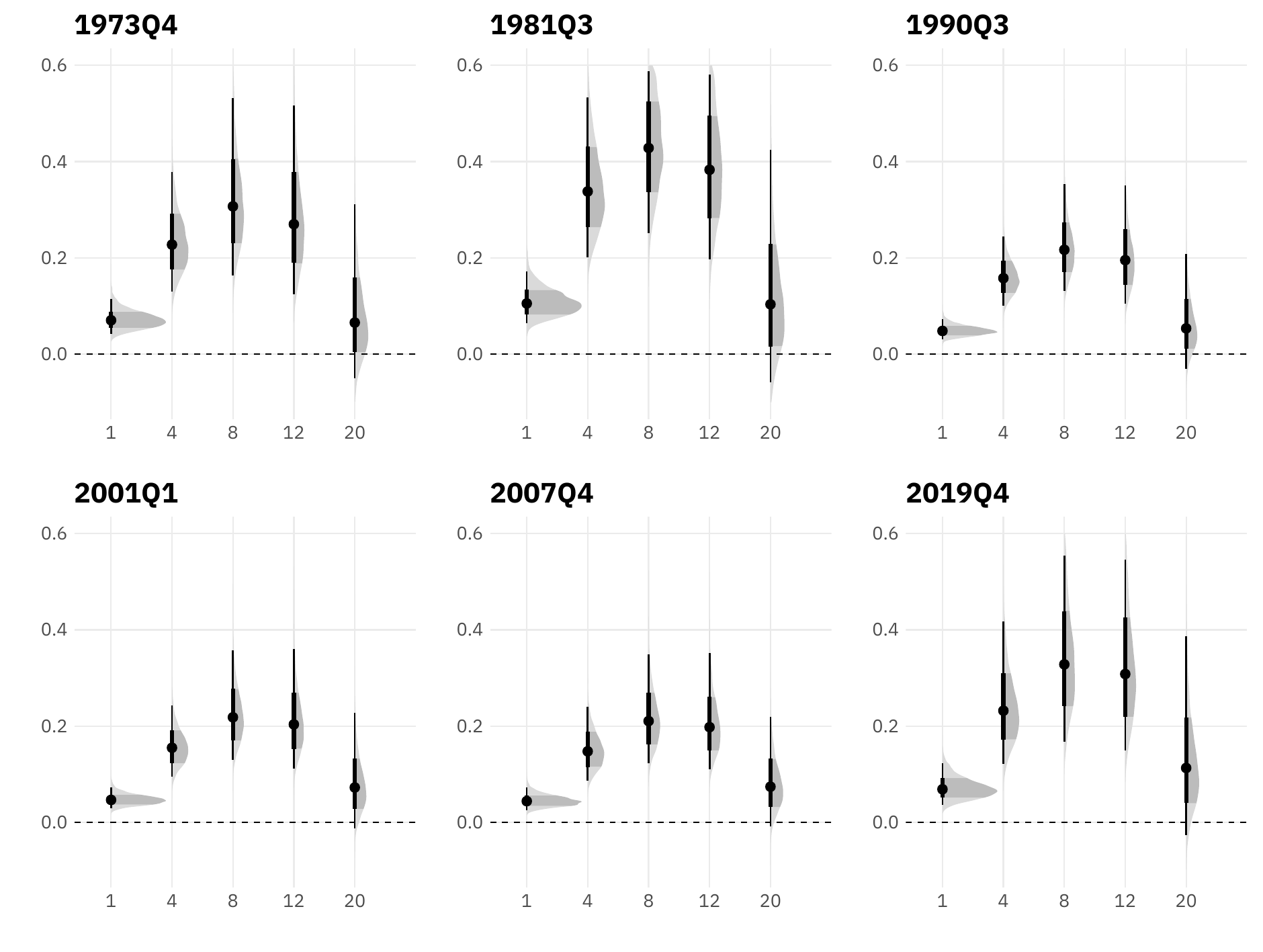}
  \caption{Posterior responses of the employment rate $(e_t)$ to a residential investment $(g_t)$ shock at different NBER business cycle peaks. Black dots denote posterior medians, and darker shaded areas the 66\% posterior density region.}
  \label{fig:e-gr-irf}
\end{figure}

Next, Figure \ref{fig:e-gr-irf} shows the posterior responses of the employment rate to a shock in residential investment. The positive feedback of accumulation on the labor market is visible across all studied cycles. The strongest responses are observed in the second and last periods, whereas the 1990s and 2000s display modest reactions. These results are consistent with theoretical and empirical works defining housing expenditures as the leading factor of the business cycle, here represented by employment increasing after a positive investment shock \citep{davis05, barbosa06, rezai13, gcri}.

\bigskip

\begin{center}
	[FIGURE \ref{fig:e-gr-irf} ABOUT HERE]
\end{center}

\bigskip

\begin{figure}[H]
  \includegraphics[width=\textwidth]{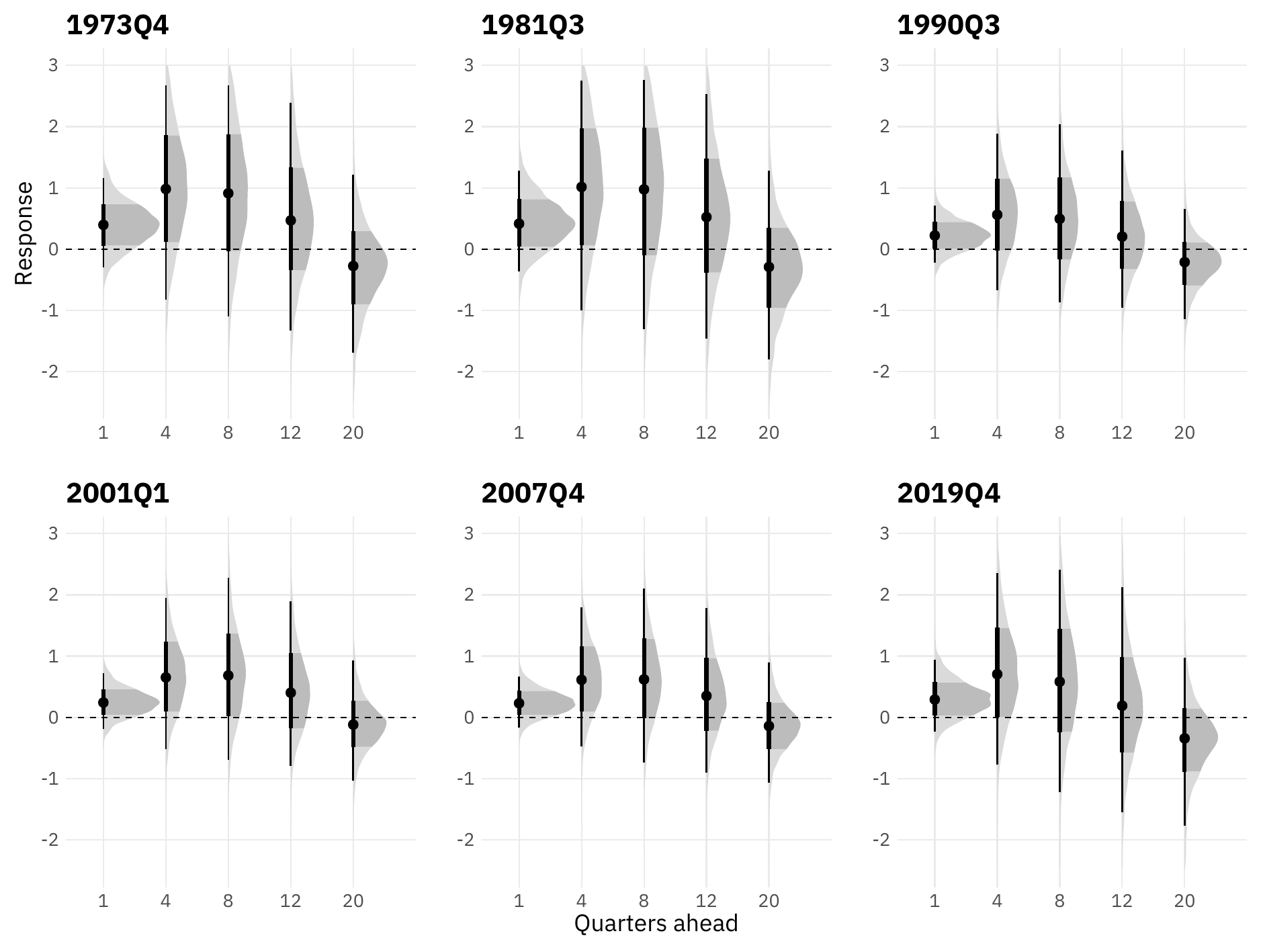}
  \caption{Posterior responses of residential investment $(g_t)$ to a shock in finance $(s_t)$ at different NBER business cycle peaks. Black dots denote posterior medians, and darker shaded areas the 66\% posterior density region.}
  \label{fig:gr-s-irf}
\end{figure}

Figure \ref{fig:gr-s-irf} illustrates the responses of residential investment to a financial shock. Overall, residential investment reacts positively to a larger term spread with a lag of one year, persisting for the subsequent quarters ahead until the last, when it becomes negative. The two cycles prior to the Great Moderation years show wider posterior distributions, with the subsequent peaks showing reduced volatility. This fact conforms with the goals of reduced volatility in economic activity envisioned by monetary policy in the second half of the 1980s.

Furthermore, these results align with works connecting real and financial cycles. A positive shock to the term spread, reflecting an expansionary monetary policy, boosts agents' expectations, as well as financial intermediaries' balance sheets and their willingness to supply credit. Consequently, interest-sensitive activity variables, such as residential investment, are highly affected by these shocks.

\bigskip

\begin{center}
	[FIGURE \ref{fig:gr-s-irf} ABOUT HERE]
\end{center}

\bigskip

\begin{figure}[H]
  \includegraphics[width=\textwidth]{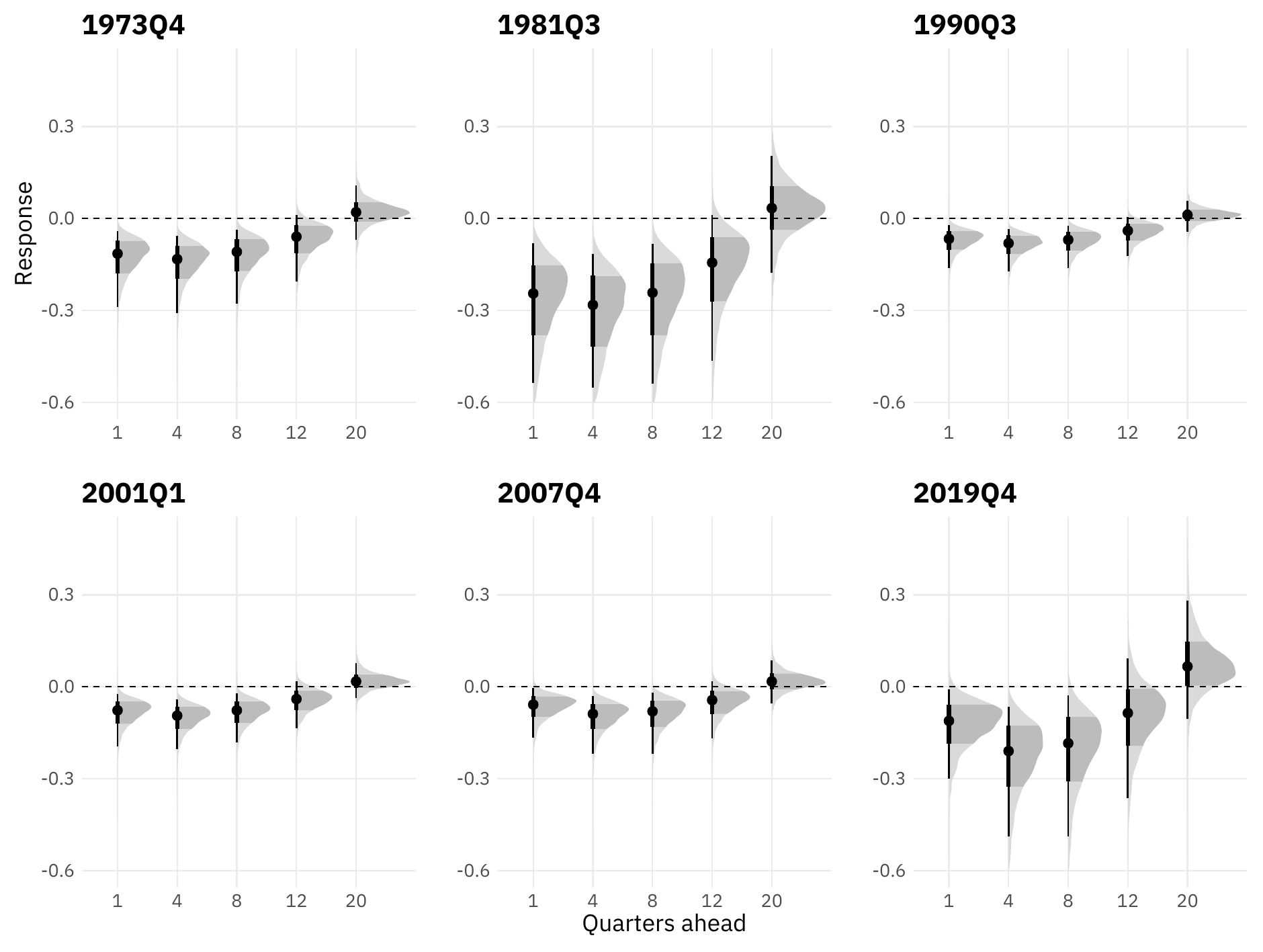}
  \caption{Posterior responses of the interest rate spread $(s_t)$ to an employment rate $(e_t)$ shock at different NBER business cycle peaks. Black dots denote posterior medians, and darker shaded areas the 66\% posterior density region.}
  \label{fig:s-e-irf}
\end{figure}

Finally, Figure \ref{fig:s-e-irf} shows posterior responses of the term spread to an employment shock. In all cycles, a tighter labor market produces negative responses of the interest rate spread. These results suggest that increased economic activity following a rise in investment---and, consequently, increased employment---trigger contractionary monetary policy movements. A reduced spread implies, in general, higher short-term interest rates, weakening expectations and reducing credit supply. It is also possible to note that the highest volatilities are shown in peaks preceding the Great Moderation years, as additional evidence of how monetary policy has acted as a means to reduce buoyancy in economic activity over the last decades.

This result aligns with Goodwinian views on the business cycle. Since the Great Moderation, contractionary monetary policy has been used to break accumulation and tighter labor market periods, under an overarching goal of controlling inflation. Higher interest rates pose a break in investment, employment, and wage demands. Once a recession is in place, financial conditions tend to ease again, setting off a new upswing in expectations, credit, and investment.

\bigskip

\begin{center}
	[FIGURE \ref{fig:s-e-irf} ABOUT HERE]
\end{center}

\bigskip

\begin{figure}[H]
  \includegraphics[width=\textwidth]{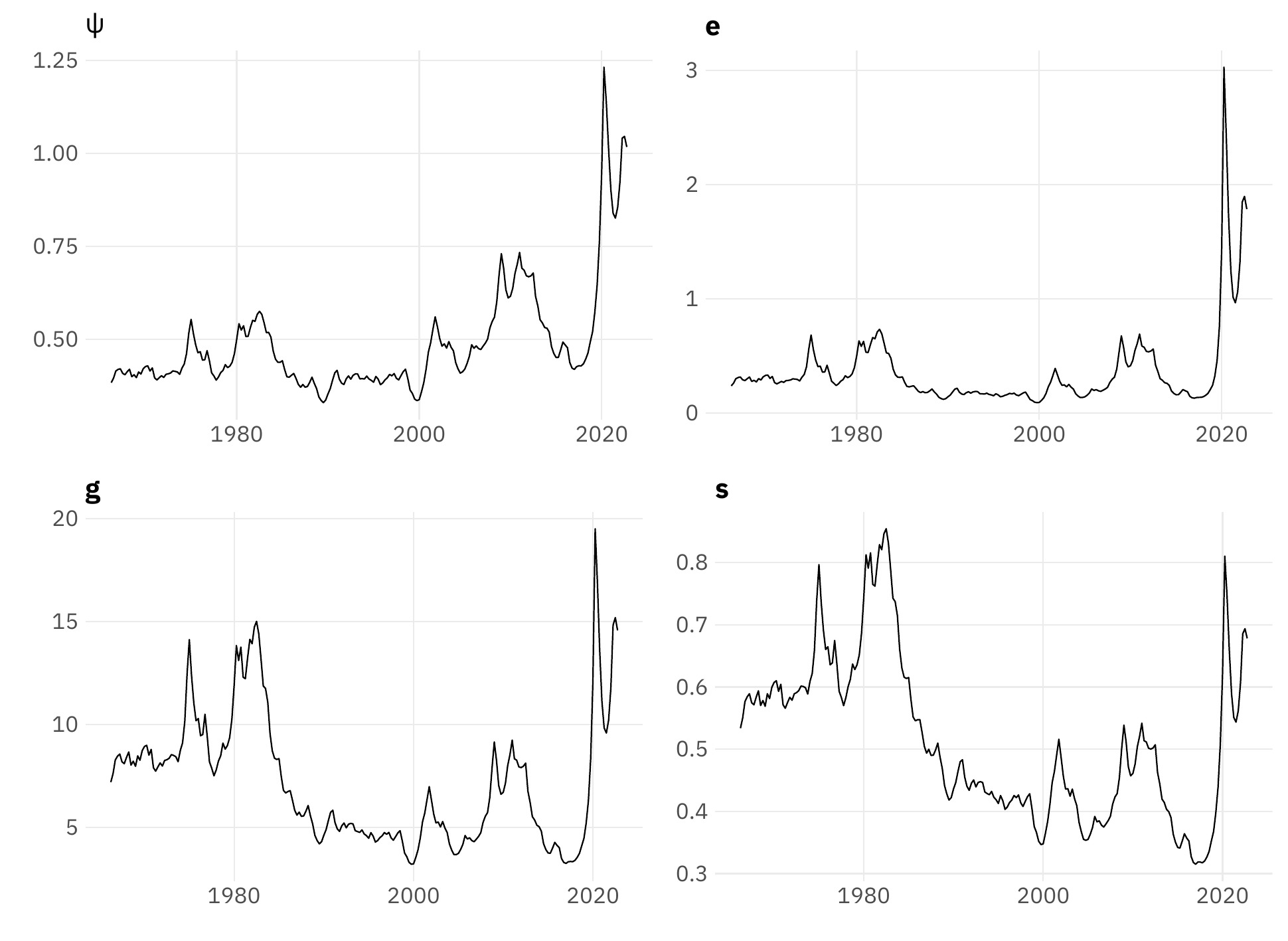}
  \caption{Volatilities of the TVP-VAR-SV model's endogenous variables.}
  \label{fig:sv}
\end{figure}

As a last visualization, Figure \ref{fig:sv} plots the standard deviations of the model residuals for each endogenous variable over time. In all four cases, clear time variations are visible. As noted in all previous impulse-response figures, years preceding the Great Moderation and as the last three years in the sample period show higher variance, whereas the Great Moderation years exhibit a reduced spread in activity, distributive, and financial measures. Such a result deviates from standard VAR procedures, where innovations are assumed to have a constant variance. Therefore, including stochastic volatility better informed the estimated model regarding exogenous variations in the data.

\bigskip

\begin{center}
	[FIGURE \ref{fig:sv} ABOUT HERE]
\end{center}

\bigskip

Models estimated for robustness purposes (including the alternative labor share and financial measures mentioned in Section \ref{sec:data}) show similar results to those shown in this section. Overall, it is possible to observe that, over shorter time horizons, the employment rate reacts negatively to a wage share shock, and the latter responds positively to a shock in the former. This pattern characterizes the profit-led/profit squeeze dynamics consistent with the Goodwinian literature on theoretical and empirical grounds. However, a weakening of both regimes, combined with the reduced volatility of economic activity envisioned by the Great Moderation policies, is a novel empirical feature provided by the present study. Finally, the most recent business-cycle peak, 2019Q4, exhibits posterior volatilities similar to those quarters preceding the neoliberal era. Incorporating most recent observations following the COVID-19 pandemic in future works will provide a more robust assessment of this apparent recovery in the Goodwin pattern.

\section{Conclusions}\label{sec:concl}

This paper proposes an empirical model relating the baseline variables of the Goodwin model---namely the employment rate and the labor share of income---with capital accumulation and its financial linkages. Following theoretical priors and descriptive statistics outlined in Sections \ref{sec:lit} and \ref{sec:data}, respectively, it identified a parsimonious, recursive Time-Varying Vector Autoregressive model with Stochastic Volatility (TVP-VAR-SV)  containing the labor share of income, the employment rate, residential investment, and the interest rate spread accounting for continuous changes in the co-movements among the variables, as well as for the presence of nonconstant variance in the structural residuals. The analysis comprehends six different U.S. business cycle peaks: 1973Q1, 1981Q3, 1990Q3, 2001Q1, 2007Q4, and 2019Q4.

The scope of this study involves three main issues. First, the demand regime experienced by U.S. economy over the post-war period. Evidence in favor of a profit-led pattern abounds in the literature. One major limitation of empirical models is that standard econometric procedures (especially VAR-type models) provide a single set of parameters that are supposed to describe an entire sample period. This way, potential structural changes experienced by an economy over time may be neglected. One recent effort  to address this issue is \cite{maldniki22}, who estimate a TVP-VAR model that finds evidence of a decline in the profit-led regime since the 1980s. The last study, however, does not include financial extensions.

Second, the distributive regime essentially consistent with the Goodwinian narrative is a profit squeeze. \cite{setterfield23} claims that this regime has weakened throughout the years following the Great Moderation due to the rise of neoliberal policies, deunionization, and business outsourcing. In short, such structural changes have institutionalized working class insecurity, thus reducing its bargaining position over income distribution. This assertion, however, is not empirically analyzed by the author. On the other hand, \cite{goldstein99} finds empirical evidence of a weakening of the profit-squeeze regime initiated in the 1970s, and aggravated after the Great Moderation years.

Third, several contributions in the Classical/Keynesian literature explore real and financial cycles together. From an empirical standpoint, \cite{rezai13} and \citet[section~5.1]{barrales21} find support for the profit-led/profit-squeeze pattern when including financial measures in their analyses. Building on literature exploring demand and distributive regimes accounting for capital accumulation as a state variable \citep{goldstein99, basu13, basugautham}, this study specifies residential investment as the channel between financial markets and real activity.

The estimation carried out in this paper deals with these three problems. First, it finds support for the results in \cite{maldniki22}: the demand regime across all studied business cycles is profit-led up to two years after an exogenous distributive shock, weakening at further horizons. In addition, posterior Impulse-Response Functions show a change in the sign of the response of the employment shock to a distributive shock. How this latter fact relates to the counter-clockwise empirical phase trajectories shown in Section \ref{sec:data} is an avenue for future research.

Secondly, the profit-squeeze mechanism has indeed declined over the last decades, as claimed by \cite{goldstein99} and \cite{setterfield23}. However, the last peak, 2019Q4, presented a more vigorous response of the wage share to labor market shocks. Whether this marks a recovery in the profit-squeeze or an isolated event, only more data will be able to shed light on this issue. Finally, residential investment reacts positively to financial shocks, represented by larger interest rate spreads. On the other hand, employment shocks induce reductions in term spreads, as an outcome of contractionary monetary policy and worsening expectations. From the 1980s until 2019, these negative reactions have become smaller in magnitude and in variability, in accordance with the goals of reducing the volatility of macroeconomic shocks following the Great Moderation.

\newpage\bibliographystyle{chicago}\bibliography{references}

\newpage

\begin{table}[H]
\centering
\resizebox{\textwidth}{!}{%
\begin{tabular}{@{}ccccccc@{}}
\hline
\textit{Variable} & \textbf{1969Q4--1973Q4} & \textbf{1973Q4--1981Q3} & \textbf{1981Q3--1990Q3} & \textbf{1990Q3--2001Q1} & \textbf{2001Q1--2007Q4} & \textbf{2007Q4--2019Q4} \\ \cmidrule(r){1-7}
$\psi$    & 1.11           & 0.826          & 1.01           & 1.00           & 1.09           & 1.04           \\
$e$      & 0.880          & 1.29           & 1.56           &  1.05          & 0.718          &                1.93		   \\
$g$      & 17.8           & 25.0           & 21.9           & 11.4           & 14.7           & 20.9           \\
$s$      & 1.07           & 1.40           & 1.09           & 1.04           & 1.18           & 0.857           \\ \cmidrule(r){1-7}
\end{tabular}%
}
\caption{Standard deviations by NBER-dated U.S. business cycles, Hamilton-filtered data.}
\label{tab:tab1}
\end{table}

\begin{table}[H]
\centering
%\resizebox{\textwidth}{!}{%
\begin{tabular}{@{}ccc@{}}
\cmidrule(r){1-3}
\textit{Variable}     & \textbf{Post-war period} & \textbf{Neoliberal era} \\ \hline
$\psi$ & 2.00                     & 2.05                    \\
$e$   & 1.49                     & 1.58                    \\
$g$   & 21.7                     & 15.8                    \\
$s$   & 1.05                     & 1.04                    \\ \hline
\end{tabular}%
%}
\caption{Standard deviations in the post-war period (1956Q1--1984Q4) and in the neoliberal era (1985Q1--2019Q4), Hamilton-filtered data.}
\label{tab:tab2}
\end{table}

\end{document}